# What the F-measure doesn't measure…

## *Features, Flaws, Fallacies and Fixes*

**David M.W. Powers,** Beijing University of Technology, China & Flinders University, Australia
*Technical Report KIT-14-001 Computer Science, Engineering & Mathematics, Flinders University*

The F-measure or F-score is one of the most commonly used "single number" measures in Information Retrieval, Natural Language Processing and Machine Learning, but it is based on a mistake, and the flawed assumptions render it unsuitable for use in most contexts! Fortunately, there are better alternatives…

## What the F-measure is!

F-measure, sometimes known as F-score or (incorrectly) the $F_1$ metric (the $\beta$=1 case of the more general measure), is a weighted harmonic mean of Recall & Precision (R & P). There are several motivations for this choice of mean. In particular, the harmonic mean is commonly appropriate when averaging rates or frequencies, but there is also a set-theoretic reason we will discuss later. The most general form, F, allows differential weighting of Recall and Precision but commonly they are given equal weight, giving rise to $F_1$ but as it is so ubiquitous this is often understood when referring to F-measure.

## Who's the F-measure for?

F-measure comes from Information Retrieval (IR) where Recall is the frequency with which relevant documents are retrieved or 'recalled' by a system, but it is known elsewhere as Sensitivity or True Positive Rate (TPR). Precision is the frequency with which retrieved documents or predictions are relevant or 'correct', and is properly a form of Accuracy, also known as Positive Predictive Value (PPV) or True Positive Accuracy (TPA). F is intended to combine these into a single measure of search 'effectiveness'.

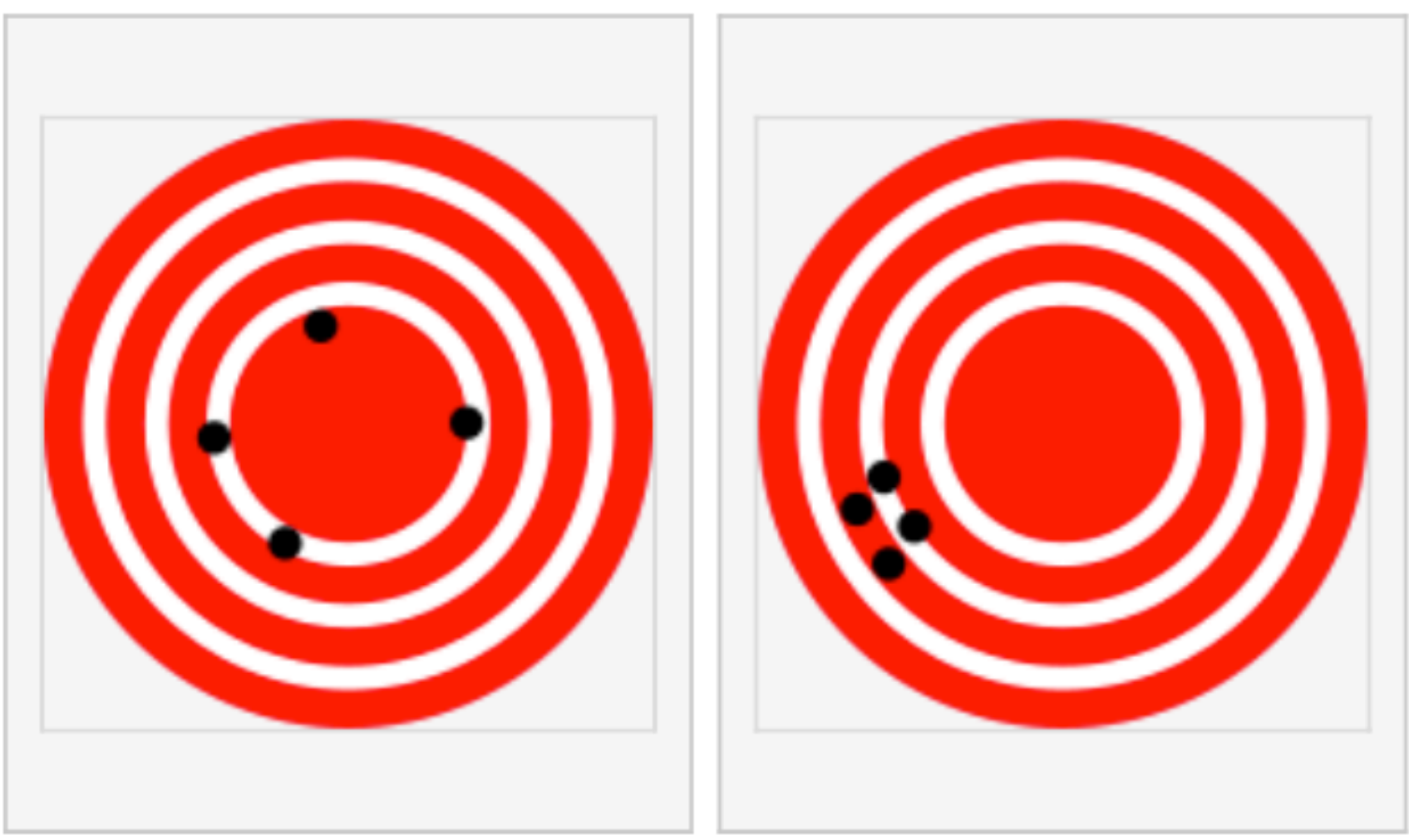

**Figure 1. Medium Accuracy with High Precision (left) or High Trueness (right) but not both together - stolen from Wikipedia (redraw)**

More generally, precision refers to the concept of consistency, or the ability to group well, while accuracy refers to how close we are to the target, and trueness refers to how close we are to a specific target on average (Figure 1)[1].

High precision and low accuracy is possible due to systematic bias. *One of the problems with Recall, Precision, F-measure and Accuracy as used in Information Retrieval is that they are easily biased*. To better understand the relationships between these measures it is useful to give their formulae in two forms, one form related to the raw counts, and one related to normalized frequencies (Equation 1 and Table 1).

These statistics are all appropriate when there is one class of items that is of interest or relevance out of a larger set of *N* items or instances. This single class of interest we call the positive class, or the real positives (RP). More generally we name multiple classes and refer to the proportion of the total each represent as its prevalence (for RPs, Prev=ρ=rp=RP/*N*). IR also assumes a retrieval mechanism that gives rise to the predicted positives, and also represents the bias of this "classifier" (Bias=π=pp=PP/*N*). In this systematic notation, we use upper case initialisms to refer to counts of items (e.g. RP), and lower case equivalents to refer to the corresponding probabilities or proportions (rp). It is also common to use Greek equivalents for probabilities in a mnemonic way (ρ).

## How's the F-measure defined?

We can now formally define Recall (R or Rec) and Precision (P or Prec) in terms of true positives (tp=TP/$N$) and the +ve prevalence (Prev) and Bias, using either their count or probability forms. Table 1 is provided to explain the notation in our equations, where we also include definitions of Accuracy (A or Acc) and F using two different means:

$$R = TP\ /\ RP = tp\ /\ rp \qquad (1)$$

$$P = TP\ /\ PP = tp\ /\ pp \qquad (2)$$

$$A = [TP+TN]/N = tp+tn \qquad (3)$$

$$F = tp/am(rp,pp) = hm(R,P) \qquad (4)$$

**Table 1. Systematic notation underlying (1-4) defining F as a harmonic mean (*hm*) and showing how this corresponds to referencing to the putative distribution given by the arithmetic mean (*am*).**

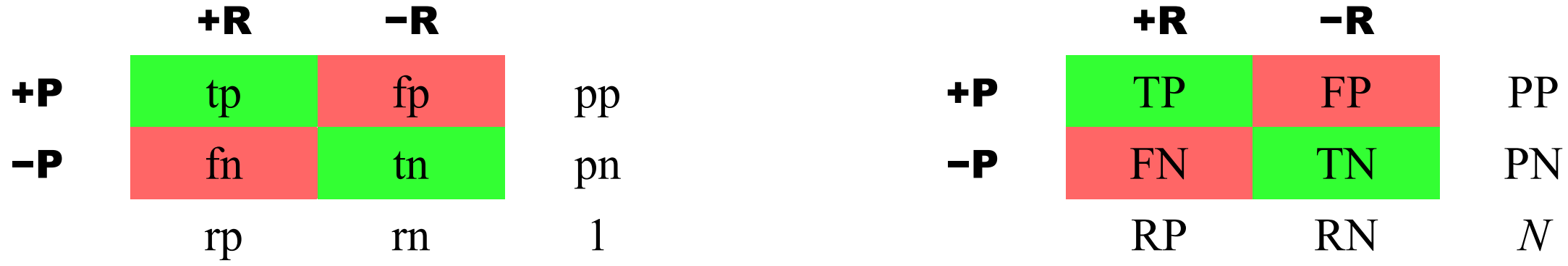

| | **+R** | **−R** | |
|---|---|---|---|
| **+P** | tp | fp | pp |
| **−P** | fn | tn | pn |
| | rp | rn | 1 |

| | **+R** | **−R** | |
|---|---|---|---|
| **+P** | TP | FP | PP |
| **−P** | FN | TN | PN |
| | RP | RN | *N* |

Note that A, R and P are themselves probabilities or proportions. Accuracy is the probability that a randomly chosen instance (positive or negative, relevant or irrelevant) will be *correct*. Recall is the probability that a randomly chosen *relevant* instance will be *predicted* (positive). Precision is the probability that a randomly chosen *predicted* instance (positive) will be *relevant*. Accuracy can also shown to be the average of Recall and Inverse Recall (viz. with positives and negatives inverted) weighted by Prevalence. Accuracy is also the average of Precision and Inverse Precision weighted by Bias.

## Why the F-measure is used!

We can now see another, set theoretic, way of looking at the definition of F-measure, and this is how it was originally defined in the equally weighted version (Figure 2)[2]. In fact, what was defined was E = 1-F which is a reinvention of the Dice semimetric[3] that was normalized by dividing by the average size of the compared sets. This is how we come to have F as a harmonic mean, as the TP intersection size is divided by the arithmetic mean of the RP and PP cardinalities. *Moreover the Dice distance measure, which failed to satisfy the triangle inequality, was a faulty reinvention of the Jaccard metric*[4] *- Jaccard normalizes more appropriately by the size of the union of the two sets (without double counting), and we note that it indeed satisfies the mathematical definition of a metric.*

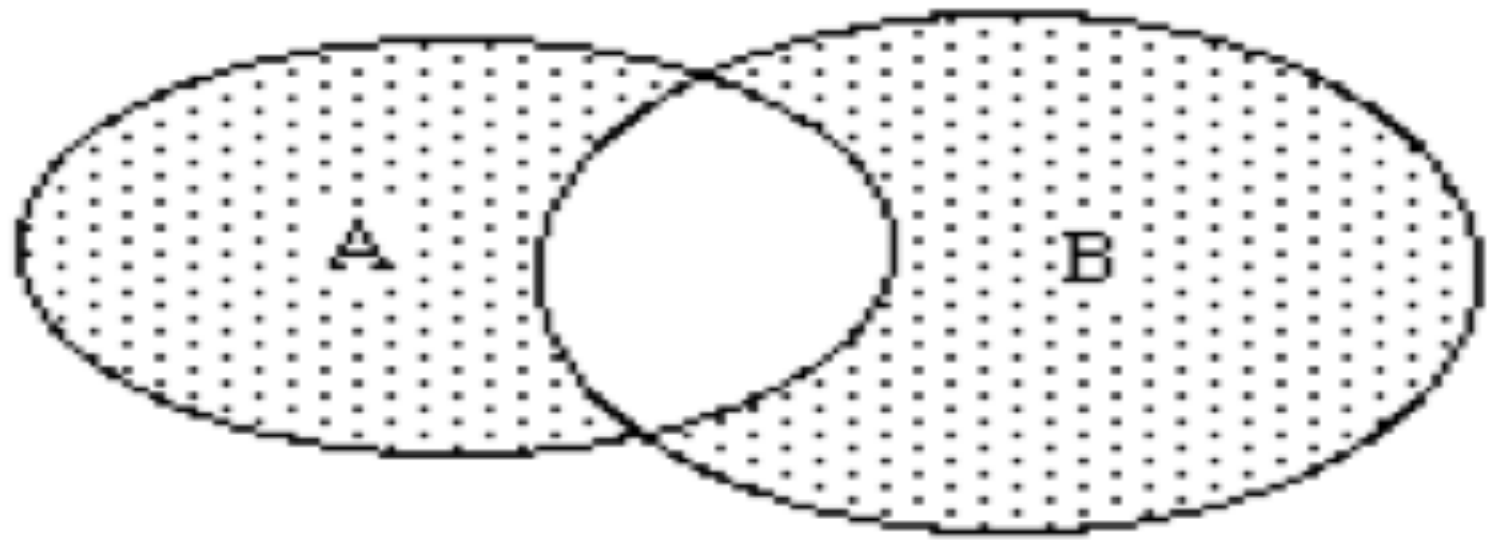

**Figure 2. Set theoretic interpretation of F-measure. Stolen from van Rijsbergen (1979) Fig 7.11 (redraw). F-measure corresponds to the number of items in the white intersection divided by the average size of the A and B groups being the Real and Predicted items.**

$F_1$ is also a reinvention of the statistical measure Positive Specific Agreement, which is designed to compare the agreement of two raters under the assumption that their ratings come from the same distribution (e.g. they are both native speakers of the same dialect with the same understanding of nouns and verbs). Thus the average of the two separate sample distributions is taken as an approximation of the underlying but unknown distribution. *This is hardly appropriate when one distribution is a real distribution based on human judgements, and the other comes from a system being evaluated.* We will however meet this assumption again when we discuss chance-corrected measures.

The F-Measure was popularized by the MUC and TREC[5] competitions based on a presentation in the 2nd edition of the early van Rijsbergen textbook on Information Retrieval[2] that recapitulated his 1974 paper. However, the function defined there was E for Effectiveness (E=1-F) and made use of a different function F. But F has stuck, and it is now virtually impossible to publish work in Information Retrieval or Natural Language Processing without including it. But that is a big mistake…

## When the F-measure is wrong!

We have already noted in passing four flaws with F-measure that emerged from theoretical considerations, and we add three further practical issues:

- F-measure (like Recall and Precision) focuses on one class only
- F-measure (like Accuracy, Recall and Precision) is biased by the majority class
- F-measure as a probability *assumes* the Real and Prediction distributions are identical
- E-measure (1-F) is not technically a metric as it does not satisfy a triangle inequality
- F-measure doesn't average meaningfully across real class or predicted label or runs
- F-measure doesn't in general take into account the True Negatives (TN)
- F-measure gives different optima from other approaches and tradeoffs.

### One-Class

The one-class issue is one of usage. If you are only interested in one class, then F-measure is not unreasonable. But note that the number of true negatives (TN) can change arbitrarily without changing F-measure (as we discuss further below), and that there is no easy way to form a macro-average across class or prediction distributions – it is assessing performance relative to a mixture of the real and prediction distribution, and if we did perform the calculations to average over that fictitious distribution, we would get Rand Accuracy (3). With Recall (1), if we macro-average weighted by the prevalence of each class, we also get Rand Accuracy (3). With Precision (2), if we macro-average weighted by the bias towards predicting each label, we also get Rand Accuracy (3).

### Bias

The bias issue is arguably the most serious, and affects both Recall and Precision as well as Accuracy. In fact, van Rijsbergen[2] reviewed techniques that dealt with bias and chance (effecively ROC and Informedness, which we discuss below). He however started with the goal of finding a way of combining Recall and Precision into a single measure, and concluded that for the purposes of Information Retrieval the additional complexity of the proposals wasn't warranted as the studies of the other methods did not conclusively show they were optimal. The E-measures that we know in complementary form as the F1-measure and the more general F-measure, were in fact special cases of a far more general approach to fitting intuitions about effectiveness and choosing the point to optimize in relation to the Recall-Precision tradeoff. But bias impacts both Recall and Precision themselves, and no form of averaging is going to get rid of it, so it is clearly suboptimal. But van Rijsbergen also notes that his general F-formula for combining them (not the F-measure we know) could be useful with measures *other than* Recall and Precision.

The damning example of bias in F-measure that brought this to our attention came from real life examples in Natural Language Processing (parsing and tagging)[6]. Here we found that a common tuning approach to increase the F-score was to dumb down the system by getting it to guess rather than use principled statistical techniques. There is a real problem here as a better system can actually get a worse F-score.

One example concerned the word 'water', which was almost always a noun (roughly 90% of the time) and much more rarely a verb (say 10%) and we can assume for simplicity that any apparent adjectival uses are in fact nominal. What the systems did was simply say water is *always* a noun. This particular form of *guessing* (always guessing noun) delivers 100% Recall, and 90% Precision, and 90% Accuracy, while any of the arithmetic, geometric or harmonic means is somewhere between the 90% and 100% mark, with arithmetic being the highest, the harmonic mean more conservative, and the arithmetically weighted average Accuracy is at the minimum.

The arithmetic, geometric and harmonic means correspond to the Lp means for $p = +1$, 0 and $-1$. The geometric mean is actually the geometric means of the paired Lp means for $p=\pm p$ and is thus is in a sense the middle mean and is usually closest to the mode and least affected by outliers. In application to Recall and Precision, this gives rise to the G-measure (which also satisfies the general form of 1-E-measure). Accuracy is based on a weighted arithmetic mean, but even though we get 0% Inverse Recall and 10% Inverse Precision, the low Inverse Bias (0%) and Prevalence (10%) mean that the weighted average doesn't help either, and this also applies for any other weighting in [0,1] that might be applied to Recall and Precision, in the most general form of the F-measure. Furthermore the limits for p=±∞ correspond to max and min, and min is the most conservative of the Lp family, but all these Lp means lie in the [90,100]% range, irrespective of weighting. Thus no choice or weighting can avoid the bias problem.

## Distribution

The distribution problem is more subtle, but when it comes down to it the correct number of positives to predict is the actual number of real positives. In the closely related *F1* and *Dice* measures, there is an explicit assumption that two different raters' class labels are drawn from the same distribution without any expectation that one is more reliable. This does not apply here, if we truly believe our 'Gold Standard' is ground truth, and that the system we are evaluating is the one that is at fault if there are discrepancies. This is clearly a problem for Machine Learning and Intelligent Systems in general.

For Information Retrieval, the situation is a little different, as there is really only one class of interest. The number of irrelevant documents is sufficiently large to be beyond our comprehension, but the number of relevant documents may also be very large, and indeed not completely known. Early in the TREC[5] competitions, the set of relevant documents was initially seeded with the documents returned by *any* of the systems, and only these were evaluated for relevance. Later *better* systems being evaluated were marked *wrong*, reducing Precision, if they returned a *relevant* document that none of the seeding systems discovered. This has been recognized by use of alternate terminology such as Coverage where only a subset of relevant documents is known. Moreover the number of documents returned is often limited by practical considerations, often to a fixed ceiling (and search engines may allow users to chose it). Furthermore, search engines will allow further blocks of hits and users will often decide to take another block two or three times, if it seems like it is worthwhile (e.g. sufficient promising leads to keep going but nothing good enough a match to stop).

Two alternatives to F-measure that became common in IR in recent years through TREC[5] are MAP (Mean Average Precision) and R-Precision. MAP is a kind of area under the Recall-Precision curve that averages over the different bias points corresponding to this "keeping going" possibility. MAP tends to correlate well with the simpler R-Precision that effectively evaluates at the point where Bias = Prevalence or PP = RP. Here of course Prevalence or RP represents the number of documents you expect which in an experimental context is the number you know you have.

Interestingly, van Rijsbergen's derivation[2] hinges on the assumption that the user will be interested in a particular tradeoff of Recall vs Precision – that is how much increase in Recall will be traded off for a decrease in Precision. This is expressed in terms of the ratio P/R and directly leads both to the weighting factor and the choice of the harmonic mean in his derivation. However, given the definitions of Recall and Precision (1&2), this corresponds to Prev/Bias ($\pi/\rho$=RP/PP). On the other hand, setting Prev=Bias corresponding to P/R=1, or any other constant value, doesn't allow for the fact that different levels of Precision mean you need different numbers of hits returned to ensure you get any desired number of relevant documents, D: D = PP/Precision. But for a particular system and user (and set of topics searched) it may be that Precision approximates a constant and the P/R ratio is thus meaningful, setting the desired Recall.

Setting P/R can also be viewed as setting a tradeoff between False Positives (FP) and False Negatives (FN), if we expand out the denominators of (1&2). In a more general Intelligent Systems context, this corresponds to setting the relative costs of False Positives and Negatives, and this is also a feature of ROC curves which trade off TPR (aka Recall) and FPR (aka Fallout = 1 – Inverse Recall).

Setting $0<P/R<\infty$ also forces $P>0$ and $R>0$, which in turn implies and requires $TP>0$. However, setting the weighting for F-measure does *not* achieve this, and does *not* solve the bias problem although it does affect the *distribution* problem. P/R ≠ 1. A choice of F-measure other than $F_1$ (β=1) does imply a bias away from the Bias = Prevalence constraint. This however considers only the mean of the distributions, and it is possible that the FP and FN errors have different distributions, both from each other and for different queries. In particular they can have different variances (as covered by van Rijsbergen in his review of ROC-related measures[2]), and this means that any setting of P/R or β other than 1 cannot be effective.

## Metricity

Whether a measure is a metric or not may seem rather academic. F is set up as a similarity measure – we want it to be 1. But E was set up as a distance or dissimilarity or error measure – we want it to be 0. This difference is similar to the difference between use of the Cosine or Correlation type measures we discuss later, which we want to be 1 for maximum similarity, versus the Euclidean distance measure, which we want to be 0. This in turn reflects a Sine function versus a Cosine function of the divergence of the vectors being compared. For purposes of graphical visualization well behaved measures are desirable, and the ability to convert sensibly into distances is desirable for similarity measures.

In terms of our intuitions, we expect things to add together in certain ways. The triangle equality is the missing element that the E and F-measures fail on. This is the idea that the shortest distance between two points is the direct line between them. For the E-measure, the sets {R} and {P}are 1/3 away from {R,P} but 1 away from each other, so it is closer to go from {R} to {P} via {R,P} than directly (where R and P here represent individual documents)!

If such measures are used for clustering it leads to very confusing, non-monotonic results.

## Averaging

We saw earlier that averaging Recall with Prevalence weighting and averaging Precision with Bias weighting both give Accuacy. This means that averaging generalizes from the two class case we have been considering to the multiclass case.

F-measure's harmonic mean essentially means we should be averaging over the putative expected (real or predicted) population, and in practice this means that macro-averaging in a principled way is not practical, and that macro-averaging based on either equal weighting or prevalence weighting, as many systems do, is not meaningful. Some systems even do different kinds of averaging in different places and get inconsistent results.

This 'apples vs pears' principle applies when we are averaging over multiple runs or multiple queries or multiple datasets as well as multiple classes. It is important always to average using weights reflecting the appropriate units. If we are talking Recall we are talking proportions relative to the actual members of a class (per real positive). If we are talking Precision we are talking proportions relative to the predictions (per predicted positive). If we are talking Accuracy we are talking about all instances (relative to $N$ if $N$ differs between datasets). If we are talking F-measure, we don't know what we are talking about! Results are relative to some fictitious intermediate distribution.

## Negatives

In Information Retrieval the negatives, the irrelevant documents, do not concern us at all. There are so many of them that the Inverse Precision and Inverse Recall are near enough to 0 to *not* be a significant factor, and thus F-measure has been argued to be a useful simplification for the sake of efficiency and comprehensibility.

Nonetheless for other Intelligent Systems, both (or multiple) classes tend to be significant for us, but neither Recall nor Precision take the TN cell of the contingency table into account. TN can be incremented from 0, to include almost all examples, without affecting the Precision or Recall, and thus without affecting F.

With this view, based on counts, we are adding new examples to the system, increasing *N*, and it is getting them all wrong.

It has however been claimed that the F-measure does in fact take TN into account if we consider RP, RN and *N* to be fixed, and *N* known. This is based on dividing by TP to recover RP and PP, and then subtracting from *N* to recover RN and PN, and thus all cells of the contingency table are determined. Technically this does *not* hold in general, specifically in the case where TP=0, and it is thus ill-conditioned as tp→0. Moreover, even when it is implicitly specified it is reflected only indirectly in the denominator in cancellation against *N* (which also does not appear explicitly).

It may be thought that we can explore the F-measure both for the positive and negative class, reversing the labels. But this can give very different answers (consider the water as noun or verb example versus the search for documents about water). In the end it comes back to how to average, and some systems do macro-average F-measure inappropriately across multiple classes (it depends on prediction as well as class).

## Tradeoffs

F-measure is about finding one number with which to compare systems and find a winner. Unfortunately this often succeeds in optimizing the wrong thing when there are more than one class of interest, because of the issues we have already discussed: F-measure is specifically seeking a trade off between Recall and Precision, but there is another pair of measures we commonly trade off in ROC: Recall (TPR) and Fallout (FPR). There have been claims PR and ROC are essentially doing the same tradeoff (see Fig. 3). In some practical contexts, this can even be true. In particular, if Bias = Prevalence (diagonals in Fig. 3) all positive measures considered become equivalent: Recall=Precision=Accuracy=F, as do the ROC and chance-corrected measures we will discuss later.

One specific claim is that 1–Precision acts as a surrogate for Fallout (1– Inverse Precision), and hence Recall-Precision is just as useful a tradeoff to consider as Recall-Fallout (ROC) since RP and RN are constant. Indeed, the relationship is quite clear as Precision = TP/PP while Fallout = FP/RN ∝ PP–TP. This however makes the tradeoff relationship quite different, although the Shannon Information conveyed by Precision, –log(Precision), interestingly gives rise to the same linear form. It also means that the areas under the curves are different and the maxima need not correspond. We can also see that taking the reciprocal for the harmonic mean of the F-measure is essentially trading off Bias (pp) against Prevalence (rp) linearly. On the other hand, in ROC, Recall versus Fallout is trading off TP against FP, normalized by constants (RP and RN) that imply differential costs for the positive and negative cases.

## Where the F-measure can be improved on!

We consider again our list of issues: one-class, bias, distribution, metricity, averaging, negatives and tradeoffs.

### One-Class

If we are dealing with more than one class, then the answer is simply to calculate Rand Accuracy directly (3) or via macro-averaging of Recall (2), as it is complex to macro-average (4) correctly, and the result would still be Rand Accuracy. Weighting F-measure simply by the size of each class (or the number of predictions of each class) enshrines a bias when these are different, and means a better result can be achieved by changing the bias towards the more prevalent classes (and some learning algorithms do this).

### Bias

There are many approaches to dealing with bias. In statistics these include regression and correlation techniques, as well as more *ad hoc* chance-correction techniques that attempt to subtract off the chance component and restore the statistic to the form of a probability. F-measure is similarly designed to retain the form of a probability, for a fictitious distribution. The tradeoff technique of Receiver Operating Characteristics (ROC) also gives rise to a method of controling for bias, and indeed there are strong relationships between all of these techniques, and we introduce them briefly now and in detail below.

Kappa: This is the *ad hoc* approach that subtracts off a chance estimate and then renormalizes to a [0,1] range by dividing by the expected error, that is the room for improvement over chance, as shown in equation (4). It was originally designed to compare human raters, but has recently been applied in Machine Learning to rate classifiers (higher kappa = better classifier), and in Classifier Fusion as a measure of Diversity (higher kappa = less diversity). The use as a measure of Diversity is closer to the original use for comparing human raters.

## Distribution

There are quite a few different versions of kappa[7] (5), the most common being Cohen Kappa, which is the one people in Machine Learning tend to know and use. But Fleiss Kappa is the one that corresponds most closely with F-measure in its distributional assumptions, and neither reflects a probability in or relative to a well defined distribution.

$$\mathrm{Kappa}_X = [\mathrm{Acc}_X - \mathrm{ExpAcc}_X] / \mathrm{ExpErr}_X \qquad (5)$$

Cohen Kappa ($\mathrm{Kappa}_C$) assumes that the two marginal distributions are independent in estimating the expected values of the contingency table due to chance – it multiples the marginal probabilities, the prevalences and biases in our context, on the assumption that they are independent distributions. Fleiss Kappa ($\mathrm{Kappa}_F$) makes the same assumption as F-measure and assumes the margins actually derive from the same distribution. Like F-measure, instead of using the actual Bias and Prevalence (and Inverse Bias and Inverse Prevalence) Fleiss replaces them with their arithmetic means before performing the same calculation as Cohen Kappa[7].

In these calculations Acc and ExpAcc (4) are the Accuracy A (3) and the expected value by adding the expected versions of tp and tn, etp and etn, while ExpErr = 1–A is the sum of the expected version of fp and fn, namely efp and efn. For Cohen Kappa we have etp = pp*rp and fp = pp*rn. For Fleiss Kappa etp = epp*erp where epp = erp = (pp+rp)/2. These can be applied not just to the two class situation we have focussed on here, but to multiple classes, and indeed to multiple raters or classifiers.

Informedness and Markedness[8] are principled Kappa forms derived from Recall and Precision and their respective inverses, but have been derived in many different ways, being also closely related to Gini and ROC.

$$\text{Informedness} = R + \text{InvR} - 1 \qquad (6)$$

$$\text{Markedness} = P + \text{InvP} - 1 \qquad (7)$$

Informedness corresponding to your edge when betting on races or speculating in the stock market, the method of marking multiple choice exams that leads to an expected mark of zero for guessing, the distance of the contingency table in ROC space above the chance line, or the Gini function of the area under the curve (AUC) subtended by that single point in ROC space[9]. Under the guise of Youden's J or DeltaP', Informedness represents a regression coefficient, and Markedness or DeltaP represents one for the opposite direction of prediction.

Informedness, $\text{Kappa}_I$ based on Recall, is the probability of an *informed* decision. Markedness, $\text{Kappa}_M$ based on Precision, is the probability of a decision variable being *marked* by the real class[7]. The G-measure associated with these derivatives of Recall and Precision, consistent with an Information theoretic instantiation of the original F function of van Rijsbergen, is Matthews Correlation[8a]. Furthermore, optimizing Informedness, Kappa or Correlation tends to give a better result than optimizing a biased measure such as F-measure or even Rand Accuracy[7c]. Optimizing ROC AUC tends to optimize Informedness but also seeks to achieve a more stable operating point – Informedness is just one component of ROC AUC[7ab].

## Metricity

Of these measures, the ones that have straightforward inversion from similarity measures (1 is best) to metric distances (0 is best) are Correlation (equivalent to Cosine measure), Informedness and Markedness (both corresponding to linear regression coefficients and interpretable as Cosines).

## Averaging

Informedness is correctly averaged weighted by Bias (since each component is the Informedness of a prediction) and Markedness is correctly averaged weighted by Prevalence (since it is the Markedness of a class). As with F-measure and G-measure, Correlation is not meaningfully macroaveraged, and if required should be calculated from the Multiclass Informedness and Markedness. If a single number is required against a Gold Standard, Multiclass Informedness is it. If you are interested in information flow in the other direction, then Markedness gives you that, and for an unsupervised comparison with the same number of classes, Correlation is recommended.  But the unsupervised case gets more complicated as the number of classes not constrained to match[10].

## Negatives

For the dichotomous case of Positive-Negative, the same result is achieved whichever class is designated positive, unlike Recall, Precision and F-measure. Informedness tells you the probability that you have made an informed decision, as opposed to guessing – the ability to bias based on Prevalence that we exploited with F-measure, for the water example… It's been eliminated!

## Tradeoffs

In regard to the relationship between Precision-Recall (PR) and Receiver Operating Characteristics (ROC) curves, the same constraint that TP, Recall and Precision are nonzero is necessary to show relationships (this means that thresholds or other parameters must be constrained to avoid this case or these points eliminated from the curve, although the zero points are traditionally shown). This has been studied comprehensively by Davis and Goadrich[11] who show that the ROC curve and PR curve for a given algorithm contain the same points, that one ROC curve dominates another in ROC space if and only if the corresponding PR curves display the same dominance. Furthermore there is an analog in PR space of the well known convex hull of ROC space, which we call an achievable PR curve. The similar linear relationships expressed by Precision Information and Fallout mean that the points in the curve correspond in a deep way, with the result that the operating points interpolated and omitted in a ROC Convex Hull correspond to the achievable and omitted points on the corresponding smoothing of the PR curve, although the interpolation is no longer linear in PR space (see Fig. 3).

Constant F-curves are indeed curves in PR space, but isocost curves in ROC space are linear and parallel, with the default cost equating the value of the full set of true positives and the full set of real negatives.

## Would the F-measure ever be the best measure?

No! There is always something better, but sometimes the error in using F-measure is small, and at times it can even vanish – just don't depend on this!

Under the constraint that Bias = Prevalence, or equivalently at the break-even point where we constrain Recall = Precision, the question of which measure to use becomes moot, In terms of IR measures, P = R = F. For the two-class ML measures, all the Kappa, Regression and Correlation variants also coincide. Furthermore optimizing any of these measures can be expected to optimize Accuracy. This constraint is thus a useful heuristic and corresponds to the default assumption that getting all the positives right is of equal value to getting all the negatives right, so you need to work equally hard on each class. Any other bias upsets this, and as extreme cases, Bias = 0 will get all the negatives right, but none of the positives (F=0), while Bias = 1 will get all the positives right, but none of the negatives (Precision=F=Prevalence, Recall=Bias=1), and the Inverse Prevalence, IPrev = 1–Prev, is thus the room for improvement or expected error for $\text{Kappa}_I$. For optimal performance we need to get both right, and Informedness and Markedness expressed in Kappa form capture this normalization for Recall and a corresponding one for Precision, with equivalence clear when Bias = Prevalence:

$K_I$ = Informedness = [R–Bias] / IPrev (8)

$K_M$ = Markedness = [P–Prev] / IBias (9)

For supervised learning or other systems where there is a 'Gold Standard', in the absence of further information about the cost or probability distribution of cases, Informedness[8] is the appropriate measure to use, and as it is the height above the chance line in ROC (Fig. 3). Receiver Operating Characteristics[9] is thus an appropriate graphical representation to use for assessing tradeoff and resilience to changes in the prevalence conditions. Moreover, ROC does also have the flexibility to explicitly manage the cost tradeoff[9] just as the general form of F-measure aims to do with its $\beta$ tradeoff parameter.

For unsupervised contexts or human-human or system-system agreement, where each side has equal status as opposed to one being a Gold Standard (which are usually rather tarnished anyway), this is the context for which Fleiss Kappa and Matthews Correlation have been developed. Matthews Correlation, the geometric mean of Informedness and Markedness, is in general an appropriate measure, providing the number of classes match[8]. To the extent that Bias tracks Prevalence, Correlation = Informedness = Markedness is the probability of information flow in *each* direction. To the extent that Bias is independent of Prevalence, the coefficient of determination, $\text{Correlation}^2$ is the joint probability of informed determination in *both* directions.

If unsupervised techniques such as clustering do not satisfy the constraint that the number of categories on one side equals the number of classes on the other, then some heuristic, e.g. a greedy approach to equate classes, can be applied to allow the use of Informedness or Correlation[8]. For this case a variety of modified or alternate techniques are available.[10]

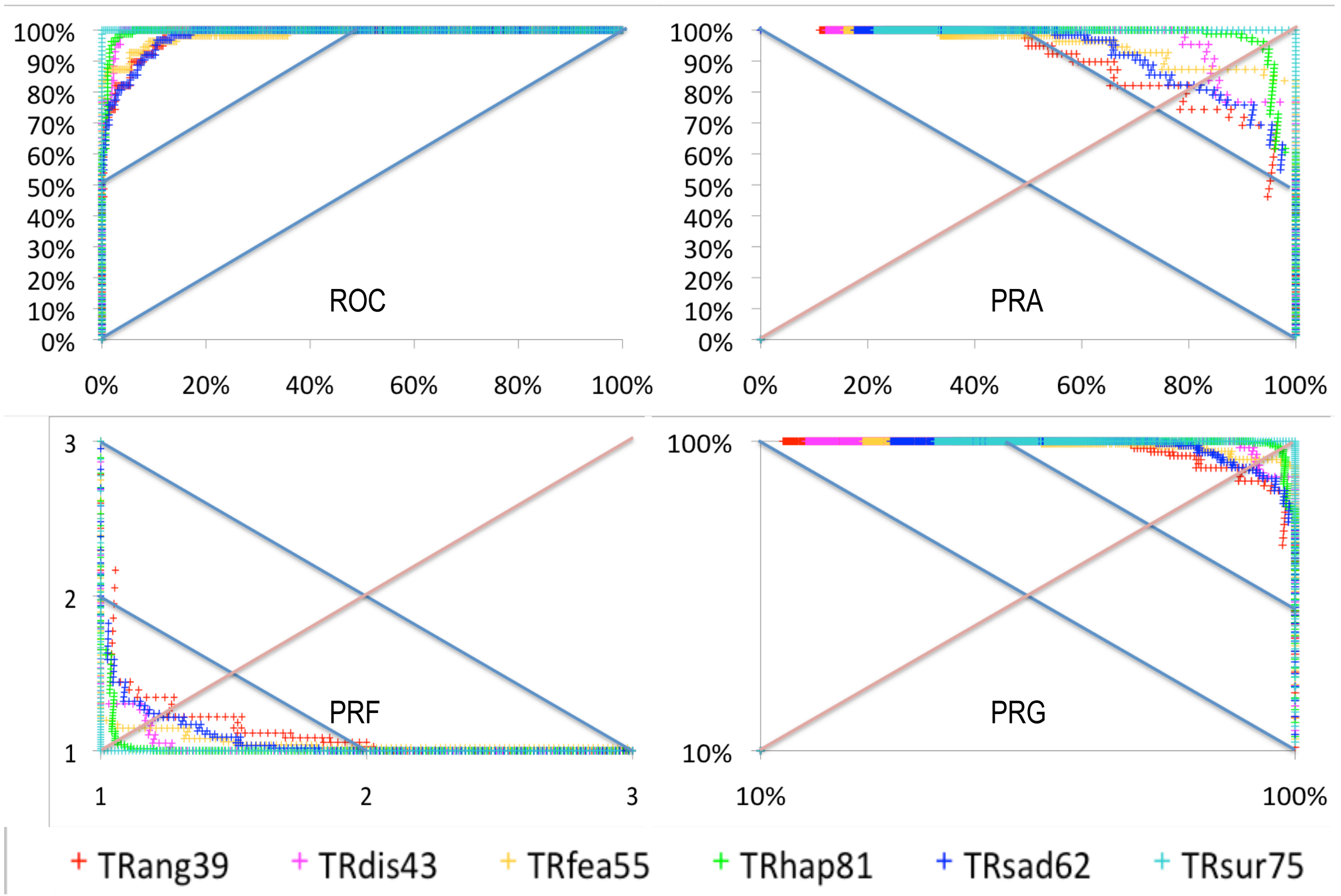

**Figure 3. Comparison of Receiver Operating Characteristics (ROC) equal Informedness isobars with Precision-Recall (PR) with Arithmetic (A), Harmonic (F) and Geometric (G) Mean isobars using Multi-Classifier Fusion for Facial Expression Recognition on the Cohn-Kanade image set[12]. True [Positive] Rate results are shown for anger, disgust, fear, happiness, sadness and surprise based on 10-fold Cross Validation with the key showing number of images per real class. Blue isobars are equal cost (default Recall=Fallout+constant in ROC) or equal score (*X-am*(Recall,Fallout)=constant in PR*X*). Brown break-even lines are also shown in the PR curves, corresponding to Bias=Prevalence and Informedness=Kappa=Correlation and Precision=Recall=Accuracy and is an appropriate constraint when variance (noise) is similar for both positives and negatives near the decision boundary).**

**The appropriate linear, logarithmic or reciprocal scalings are used to permit isobars to be linear: X-axis is Fallout (FRemo) for ROC and Precision (PRemo) for PRA (linear scale) and PRG (log scale); and Y-axis is reciprocal of Recall (TRemo) and X-axis reciprocal of Precision (PRemo) for PRF. Note that we 6 classes, so 1/6 or 16.7% is chance Recall and Precision, without distributional data. Therefore for PRF, the axes representing reciprocal of Recall and Precision truncate at 2x this level, and K=6F represents that we are doing Kx better relative to this naive 1/6 baseline chance level. The average angle of transition from predicting positives to predicting negatives at the threshold approximates that of the default weighting, for all of the measures, across all six of the curves. The ROC curve is smoother and thus tuning for costs seems more appropriate than for any PRX. The sharper the elbow, the less tuning the β for F or the costs for ROC will affect the optimum.**

## References (seminal and culminal)